\documentclass[%reprint,
preprint,
 amsmath,amssymb,
 aps,
]{revtex4-2}

\usepackage{graphicx}% Include figure files
\usepackage{dcolumn}% Align table columns on decimal point
\usepackage{bm}% bold math
\usepackage{MnSymbol}
\expandafter\let\csname equation*\endcsname=\relax 
\expandafter\let\csname endequation*\endcsname=\relax 
\usepackage{amsmath}
\usepackage{bm}

\begin{document}

\title{En route to nanoscopic quantum optical imaging: counting emitters with photon-number-resolving detectors}%

\author{Shuo Li$^1$, Wenchao Li$^2$, Vladislav V. Yakovlev$^3$, Allison Kealy$^2$, Andrew D. Greentree$^{1,}$}

\email{andrew.greentree@rmit.edu.au}

\address{1 ARC Centre of Excellence for Nanoscale BioPhotonics, RMIT University, Melbourne, VIC 3001, Australia\\ 
2 School of Science, RMIT University, Melbourne, VIC 3001, Australia\\ 
3 Department of Biomedical Engineering, Texas A\&M University, College Station, TX 77843, USA\\
* andrew.greentree@rmit.edu.au}


\begin{abstract}
Fundamental understanding of biological pathways requires minimally invasive nanoscopic optical resolution imaging. Many approaches to high-resolution imaging rely on localization of single emitters, such as fluorescent molecule or quantum dot. Exact determination of the number of such emitters in an imaging volume is essential for a number of applications; however, in a commonly employed intensity-based microscopy it is not possible to distinguish individual emitters without initial knowledge of system parameters.  Here we explore how quantum measurements of the emitted photons using photon number resolving detectors can be used to address this challenging task. In the proposed new approach, the problem of counting emitters reduces to the task of determining differences between the emitted photons and the Poisson limit. We show that quantum measurements of the number of photons emitted from an ensemble of emitters enable the determination of both the number of emitters and the probability of emission. This method can be applied for any type of emitters, including Raman and infrared emitters, which makes it a truly universal way to achieve super-resolution optical imaging. The scaling laws of this new approach are presented by the Cramer-Rao Lower Bounds and define the extent this technique can be used for quantum optical imaging with nanoscopic resolution. 
\end{abstract}

\maketitle

%%%%%%%%%%%%%%%%%%%%%%%%%%  body  %%%%%%%%%%%%%%%%%%%%%%%%%%
\section{Introduction}

The 2014 Nobel Prize in Chemistry was awarded for development of methods of super-resolution optical imaging, which, in particular, relied on single optical emitters imaging and localization \cite{Betzig2006science,Dickson1997nature}. Despite a tremendous progress of nanoscopic fluorescence-based imaging, which has made possible through those pioneering work, identification of single emitters remains to be a challenge \cite{Mortensen2010natmeth} and often relies on ultra-bright emission, which is not always affordable for biological systems, and some prior knowledge of the system, which is often not available. Thus, it would be highly desirable to be able to characterize individual emitters and quantify their presence in any given imaging volume. This can be generalized to a broader fundamental problem of counting the number of emitters in a sample from optically collected data, which has significant implications beyond the commonly used fluorescence imaging. For example, %… 
%and so on, as in the original text.
%
%
%
%The determination of the number of emitters in a sample from optical data is a fundamental and important practical probe.  For example,  in the case of Raman imaging, molecules can be very close to each other, forming a relatively large ensemble within an  optical diffraction limit. This leads to the question: how can we determine the number of emitters, for example 
is it possible to determine the number of emitters contributing to a fluorescence or Raman signal, on the basis of imaging data alone?  Furthermore, if it is possible, what is the limit to which we may determine the number of emitters?  In particular we want to be able to determine the number of molecules, which might be in the range of 10, 100, or 1,000.  

\begin{figure}
    \centering
    \includegraphics[width=\textwidth]{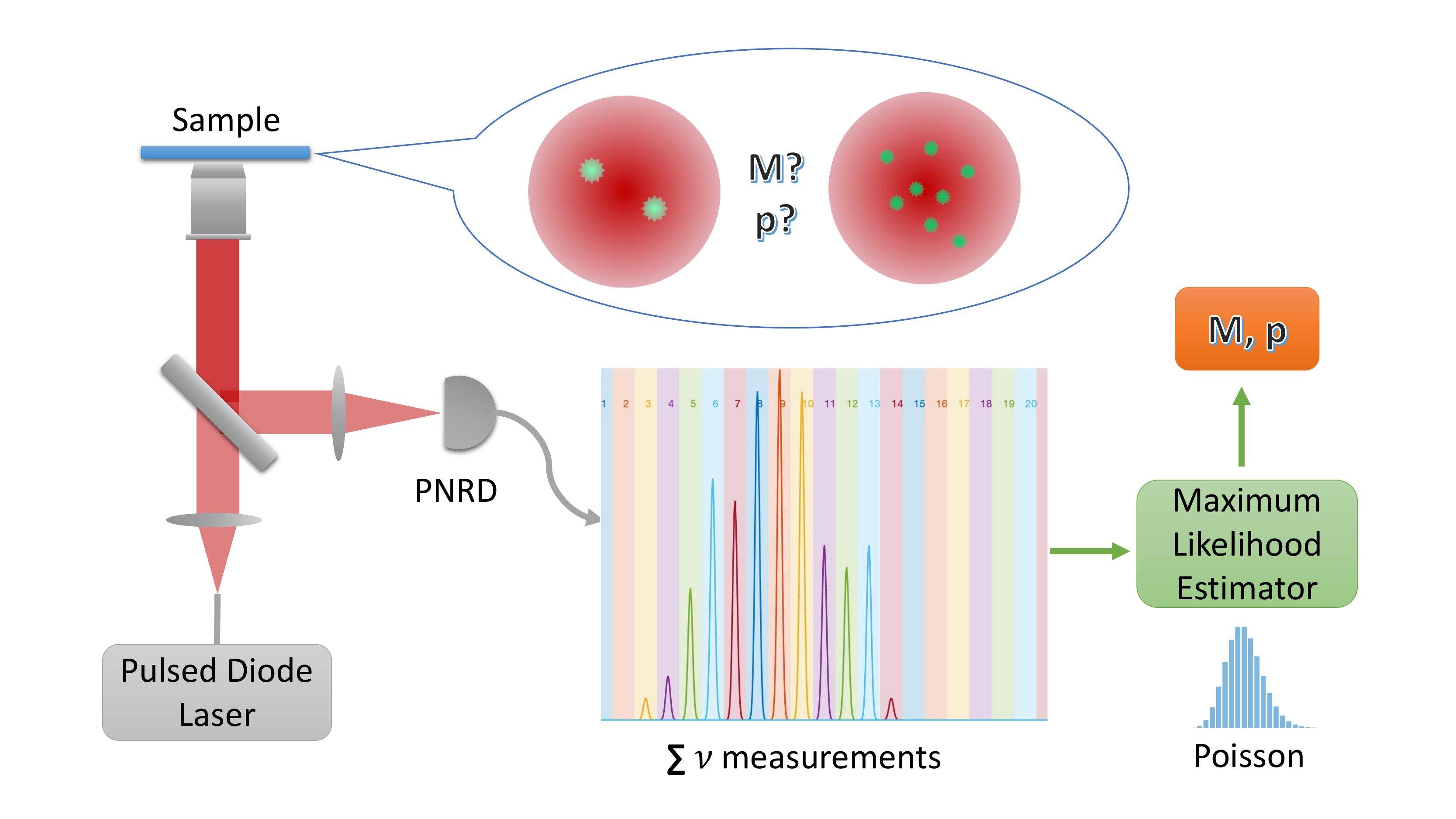}
    \caption{Schematic showing the concept to determine the  number of emitters $M$ in one sample with unknown detection probability $P$. A pulsed laser excites the sample and emitted photons are measured using a the photon-number-resolving detector, PNRD.  a histogram of the number of photons in each collected pulse in analysed using a Maximum Likelihood Estimator, which enables the determination of both $M$ and $P$.  Such discrimination is not possible using only classical intensity measurements.}
    \label{fig:schematic}
\end{figure}

The problem with determination of the number of emitters from intensity is fundamental.  For a classical fluorescence measurement, we observe an intensity, which we may express as $I = M I_0$, where $I$ is the total intensity, $M$ the number of emitters and $I_0$ is the intensity of each emitter, where we assume each emitter has the same emission intensity, for simplicity.  However, without \emph{a priori} knowledge of $I_0$ or $M$, it is not possible to discriminate between cases where there are more dimmer emitters, or fewer brighter emitters.  Additional practical confouding issues include uncertainty in the probability of collecting light emitted from the emitters, which may vary due to experimental conditions.

Quantum mechanically, however, there is the prospect for distinguishing different emitter configurations.  It is known that photon anti-bunching signals (Hanbury Brown and Twiss experiment) can be used to distinguish the number of emitters \cite{Monticone2014prl}.  This technique is typically used when the number of emitters is few (e.g. to determine the difference between 1, 2 or 3 emitters) \cite{Worboys2020pra, Davin2021arxiv, Chen2017nphoton, Schwartz2013nanoletter}. The Hanbury Brown and Twiss (HBT) experiment, in its simplest form, uses two SPDs that sample the same optical field of view via a beamsplitter \cite{Stevens2013book}.  By observing the signal in coincidence, some information about the number of emitters can be obtained, which leads to the well-known result for the background-free, equal brightness HBT signal at coincidence:
\begin{align}
    g^{(2)}(0) = 1 - \frac{1}{M}.
\end{align}
One way to improve determination of the number of emitters is to increase the number of SPDs. This approach leads to considerable complexity due to the increased number of beamsplitters and coincidence electronics that is required \cite{Steven2014oe}.  

Measurement of photon number has traditionally been a difficult task.  Originally, this task was performed using single photon detectors (SPD) such as photomultiplier tubes, and later avalanche photodiodes. Such devices permit a binary measurement of the number of photons: they measure either 0 photons, or more than 0 photons, but a single device typically does not allow for a  more sophisticated determination of the number of photons. Alternatively, new generations of photon number resolving (PNR) detectors are becoming available. PNR detectors have the ability to perform a direct projective measurement of the number of photons in a pulse of light. Compared with non-PNR detection, PNR detection can provide more information about noise and receiver imperfections \cite{Becerra2015natphotonics}. Several techniques have been applied in realising photon number resolving \cite{Provaznik2020oe, Thekkadath2020thesis}, including multiplexed APD \cite{Kardynal2008nphoton}, CMOS image sensors \cite{Ma2017optica}, superconductor nanowire \cite{Cahall2017optica}, and superconducting transition-edge sensor (TES) \cite{Schmidt2018LowTemPhys}. Additionally, there are  multipixel photon counters (MPPC) that have the ability to distinguish from one up to 10 photons \cite{Kalashnikov2011oe}. Superconducting transition-edge sensors have recently been reported to resolve photon numbers up to 16 with the efficiency of over 90\% \cite{Morais2020arXiv}. A study has reported a 24-pixel PNR detector based on superconducting nanowires that achieves the detection of $n=0-24$ photons \cite{Mattioli2016oe}. Given the increase in technology it is is expected that this upper limit will soon be exceeded, and the availability of such detectors will become more widespread.  It is therefore timely to see the effects that such detectors will have on the determination of the number of emitters in an unknown sample.  

Here we show that photon number resolving measurements enable the determination of emitter number more generally. The schematic is shown in Fig.\ref{fig:schematic}.  We theoretically determine the photon number probability distribution for $M$ emitters, with photon detection probability $p$.  On the basis of this, we show maximum likelihood estimation and the Cramer-Rao lower bound for the simultaneous determination of both the number of emitters and the probability of detection.  This analysis enables us to provide scaling laws for the number of experiments required to distinguish between different configurations.  

This paper is organised as follows: We first discuss the photon statistics from an ensemble of $M$ classically identical emitters (ie emitters with the same emission probability in the same field of view with the same emission properties such as polarisation and wavelength, although we stress that the emitters are assumed to be not quantum indistinguishable).  We then show the maximum likelihood determination of the number of emitters and photon detection probability for particular cases, as a function of the number of experiments.  Lastly we present the Cramer-Rao lower bound for the scaling.

%%%%%%%%%%%%%%%%%%%%%%%%%%%%%%%%%%%%%%%%%%%%%%%%%%%%%%%%%%%%%%%%
\section{Photon number resolving detection probabilities}

We are concerned with the problem of simultaneously determining the number of emitters, and the collection probability for a number of emitters.  We consider an experimental configuration where $M$ (unknown) emitters are excited by a short pulse laser, and the fluorescence signal collected confocally.  Each emitter is assumed to emit no more than one photon per excitation pulse, and we assume that the probability of detecting a photon from each emitter in that pulse is $p$.  The photon resolving detector performs a projective measurement in the photon number basis, and we may write down the binomially distributed probability of detecting $N$ photons from the $M$ emitters as 

\begin{align}
\mathcal{P}(N|M,p) = \frac{M!}{\left(M-N\right)! N!} p^N \left(1 - p\right)^{M-N}. \label{eq:P(N)}
\end{align}

We can explore Eq.~\ref{eq:P(N)} in various limits, however to address the original problem, the clearest case to consider is where we have a known (measured) brightness, but where the actual number of emitters and their probability of emission is unknown.  Therefore,  we set $\lambda = M p$, so that $\lambda$ is the expected number of photons emitted per experiment, where each experiment is a Bernouilli trial.  Note that although $N$ is quantised, $\lambda$ is not.  

\begin{figure}
    \centering
    \includegraphics[trim=0cm 5cm 0cm 5cm, width=\textwidth]{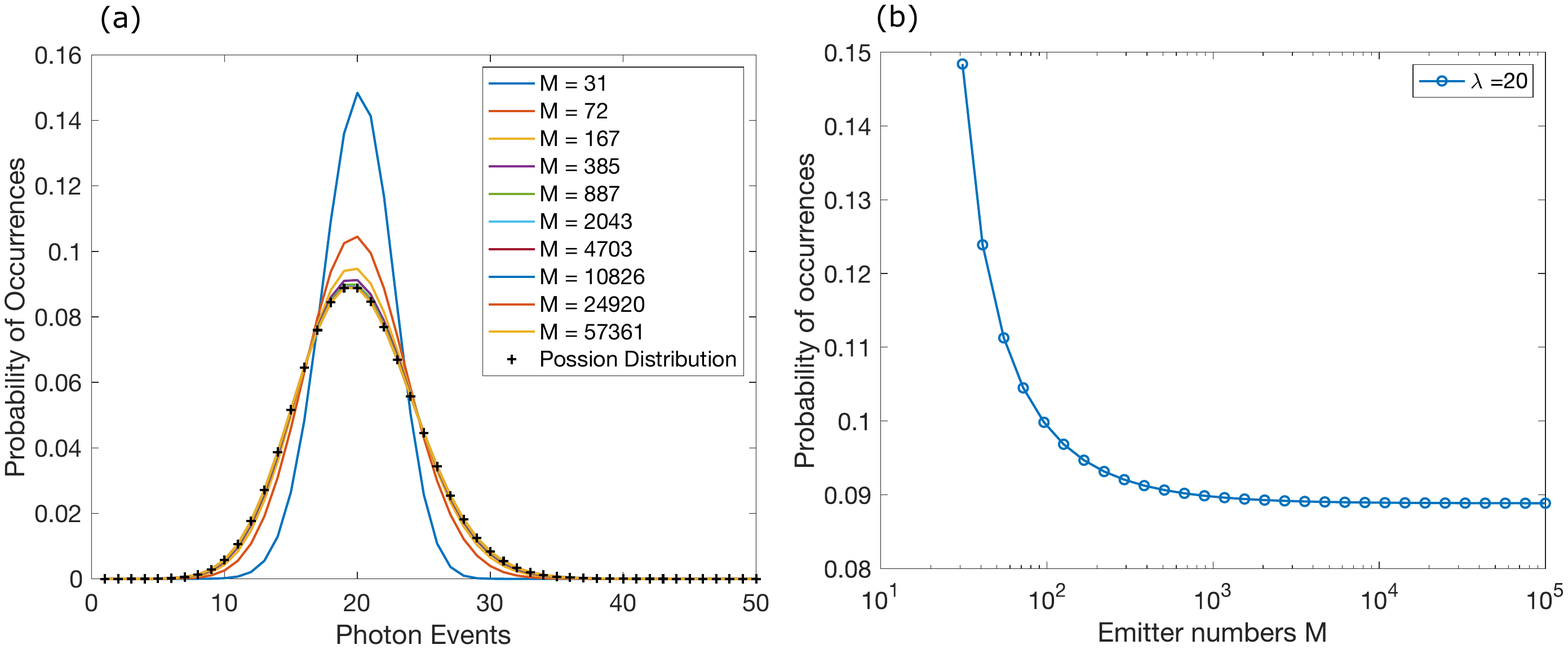}
\caption{(a) Series of curves showing the distribution of probability distribution function for the number of photons, $N$, for different $M$ and $p$ such that $\lambda = Mp = 20$. The dotted line shows the limit for a Poisson distribution.  As $M$ increases the peak broadens, and approaches the Poisson limit. By measuring the distribution, the number of emitters should be distinguishable, however it is important to stress that the differences are small, and noise will make sure determination difficult. (b) The peak of the probability distribution, at $N = 
\lambda$, is the point that shows the largest dependence on number of emitters however as this curve shows, even variation of $M$ over three orders of magnitude only leads to a change in the probability of $N=20$ photon events from  $\mathcal{P}(N = 20| M = 30, \lambda = 20) = 15.3\%$ to $\mathcal{P}(N = 20| M = 10^5, \lambda = 20) = 8.88\%$}
\label{fig:poissonDistri}
\end{figure}

Eq.~\ref{eq:P(N)} in terms of $\lambda$ becomes
\begin{align}
    \mathcal{P}(N|\lambda,M) = \frac{M!}{\left(M-N\right)! N!} \left(\frac{\lambda}{M}\right)^N \left(1 - \frac{\lambda}{M}\right)^{M-N}.
\end{align}
This result should be compared with the standard Poisson distribution, which is expected in the limit $M\rightarrow \infty$
\begin{align}
    \lim_{M\rightarrow\infty}\mathcal{P}(N|M,p) = \frac{\lambda^N e^{-\lambda}}{N!}.
\end{align}
Analytical results for this are shown in Fig.~\ref{fig:poissonDistri}. Fig.~\ref{fig:poissonDistri}(a) shows the probability of obtaining $N$ photons for different $M$.  As shown in Fig.~\ref{fig:poissonDistri}(a), the greatest change in $\mathcal{P}(N)$ occurs at $N\approx \lambda$, although it is clear that the \emph{entire} distribution provides information about $M$.  Hence it is important that any photon resolving detector should be at least able to detect $\lambda$ photons for maximum ability to determine $M$.

\begin{figure}[htb]
    \centering
    \includegraphics[width=0.5\textwidth]{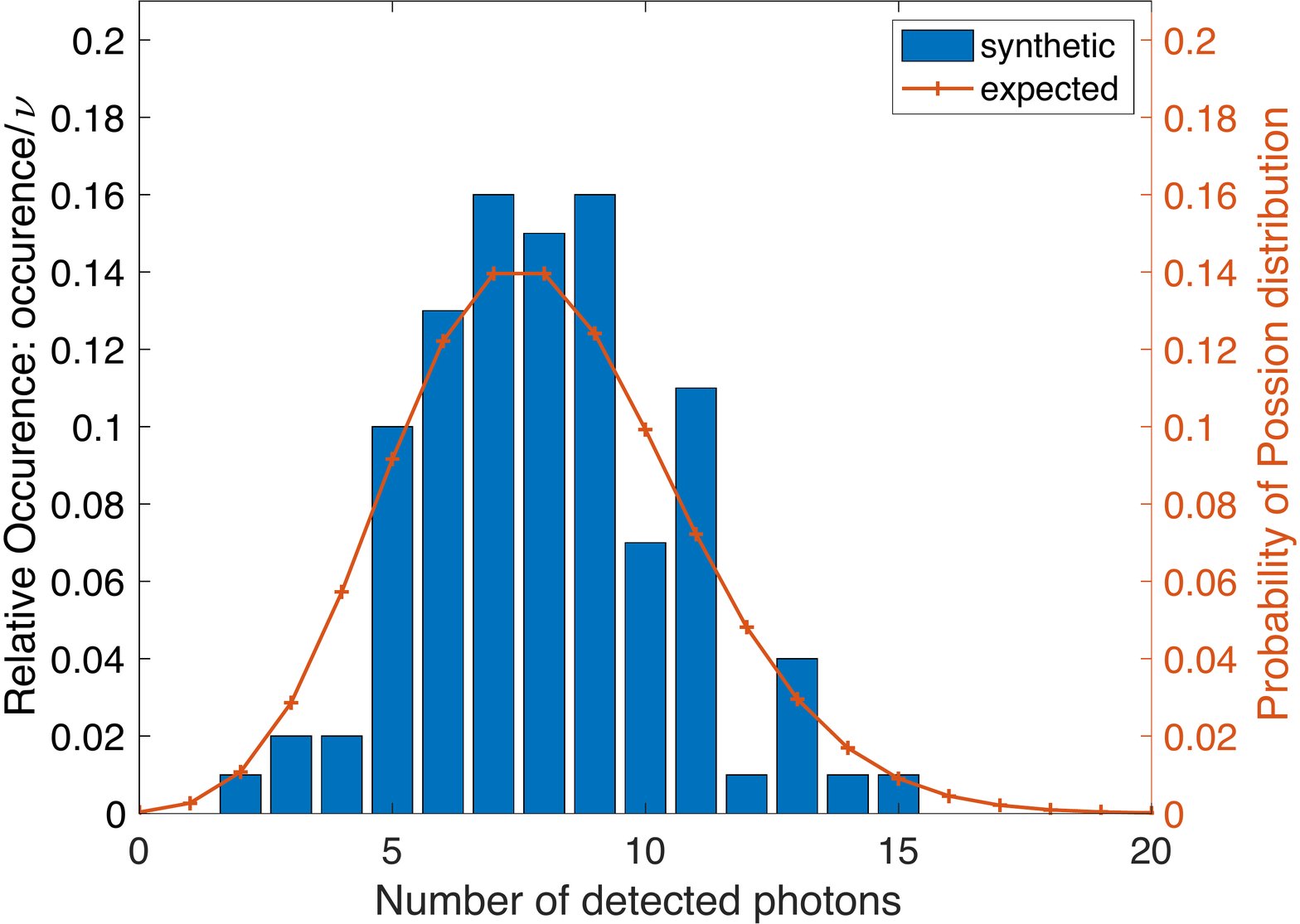}
    \caption{Plot showing relative frequency data (bars) and expected (ideal) probability of detecting a number of photons given $M=40$ emitters, each with a detection probability of $p=0.2$.  The synthetic data was generated on the basis of 100 experiments and assumes no noise other than random fluctuations provided by Eq.~\ref{eq:P(N)}.
    }
    \label{fig:BarFig} 
\end{figure}

To explore the determination of both $M$ and $p$, we begin by generating synthetic data obtained by sampling Eq.~\ref{eq:P(N)} for a finite number of numerical experiments.  This yields a histogram of events, such as that shown in Fig.~\ref{fig:BarFig}.  This data was generated on the basis of $\nu = 100$ experiments, with $M=40$ atoms and probability of detection $p=0.2$.  Also shown is the probability distribution function, $\mathcal{P}(N)$ under the same circumstances. As the number of experiments increases, the synthetic data and probability distribution function should converge.  

To address the issue of how well we can determine the model parameters using the data, we turn to maximum likelihood estimation.  For a given set of data $\bm{N} = \left(N_1, N_2, ...N_{\nu}\right)$, where the $N_i$ correspond to the number of photons resolved in experiment $i$, which are independent and identically distributed with probability mass function $\mathcal{P}(N_i|\bm{\theta})$, where $\bm{\theta} = \left(M,p\right)\in\mathbb{Z}^+\bigtimes[0,1]$ is the vector of parameters to be estimated. Furthermore, let $\bm{\theta}_0$ be the ground truth and $L(\bm{\theta}|N_i)$ the associated likelihood function of $\bm{\theta}$ given data $N_i$. 

Then we can write the joint log likelihood function as follows:
\begin{align}
\ell\left(\bm{\theta}\big|\mathbf{N}\right) = \log L\left(\bm{\theta}\big|\mathbf{N}\right)=\sum_{i = 1}^{\nu}\log L\left(\bm{\theta}\big|{N}_i\right),
\end{align}
where $\nu$ is the number of experiments.  

Accordingly, the MLE of $\bm{\theta}$ is given by
\begin{align}
\hat{\bm{\theta}}=\mathop{\arg \max}\limits_{\bm{\theta}\in\mathbb{Z}^+\bigtimes [0,1]} \ell\left(\bm{\theta}\big|\mathbf{N}\right).
\end{align}
As $\nu$ increases, from the consistency of MLE \cite{rao1973linear}, we expect that the $\hat{\bm{\theta}}$ approaches $\bm{\theta}_0$. 

It is easier to determine sample brightness than the number of emitters.  This accords with our classical intuition, namely that on the basis of intensity-only measurements it should be \emph{only} possible to determine the mean brightness, and \emph{impossible} to determine the number of emitters (ie few bright emitters should be indistinguishable from many dim emitters).  It is therefore useful to transform our parameters from $\bm{\theta}=(M,p)$ to $\bm{\beta}=(\lambda, \xi)$ where $\xi = M/p$.  With this parameterisation, the probability distribution function becomes
\begin{align}
\mathcal{P}(N|\bm{\beta})=\mathcal{P}(N|\lambda,\xi)=\frac{(\sqrt{\lambda\xi})!}{(\sqrt{\lambda\xi}-N)!N!} \left(\frac{\lambda}{\xi}\right)^{N}\left[1-\left(\frac{\lambda}{\xi}\right)\right]^{\sqrt{\lambda\xi}-N}\label{pdf2}
\end{align}

%%%%%%%%%%%%%%%%%%%%%%%%%%%%%%%%%%%%%%%%%%%%%%%%%%%%%%%%%%%%%%%%
\section{Uncertainty of the estimates: Cramer Rao lower bound}

To obtain the scaling laws for estimating $M$, and $p$, we now proceed to calculate the Cramer-Rao lower bound (CRLB). The Cramer-Rao Lower Bound (CRLB) gives a lower estimate for the variance of an unbiased estimator. The Fisher Information Matrix (FIM)\cite{Nishiyama2019arxiv,Ly2017jmp} is required to calculate the CRLB.  To do this, we need to find the derivative of (\ref{pdf2}) w.r.t $\lambda$ and $\xi$. However the likelihood function $L(\bm{\beta}|N)$ is not differentiable since $\sqrt{\lambda\xi}\in\mathbb{Z}^+$. To implement the derivative we use the $x!=x\Gamma(x)$ to transfer $(\sqrt{\lambda\xi})!$ into a continuous function with respect to $\lambda$ and $\xi$. Additionally, we have $\left[x\Gamma(x)\right]'=\Gamma(x)+x\Gamma(x)\psi(x),$
where $\psi(\cdot)$ is digamma function.

Let $\bar{L}(\bm{\beta}|N)$ be the approximated likelihood function (after replacing the factorial term associated to $\lambda$ and $\xi$ by the interpolation function), then 
\begin{align}
 &\frac{\partial \bar{L}(\bm{\beta}|N)}{\partial\lambda}\notag\\ 
=&\sqrt{\frac{\xi}{\lambda  }}\alpha_1\left\{\xi  \lambda ^2+N^2 \sqrt{\xi  \lambda }-N \left[\left(\sqrt{\xi  \lambda }-1+\lambda \right) \sqrt{\xi  \lambda }+\lambda  \right]+\left(\lambda -\sqrt{\xi  \lambda }\right)\alpha_2\right\}\\
&\frac{\partial \bar{L}(\bm{\beta}|N)}{\partial\xi}\notag\\ 
=&\lambda\alpha_1\left[-\lambda  \sqrt{\xi  \lambda }-N^2+N \left(-\sqrt{\frac{\lambda }{\xi }}+\sqrt{\xi  \lambda }+\lambda +1\right)+\left(\sqrt{\frac{\lambda }{\xi }}-1\right)\alpha_2\right]
\end{align}
where
\begin{align}
    \alpha_1&=\frac{\Gamma \left(\sqrt{\xi  \lambda }\right) \left(\frac{\lambda }{\xi }\right)^{N/2} \left(1-\sqrt{\frac{\lambda }{\xi }}\right)^{\sqrt{\xi  \lambda }-N-1}}{2 N! \sqrt{\xi  \lambda } \left(\sqrt{\xi  \lambda }-N\right)^2 \Gamma \left(\sqrt{\xi  \lambda }-N\right)}\\
    \alpha_2&= \left(\xi  \lambda -N \sqrt{\xi  \lambda }\right) \left[\log \left(1-\sqrt{\frac{\lambda }{\xi }}\right)+\psi\left(\sqrt{\xi  \lambda }\right)-\psi \left(\sqrt{\xi  \lambda }-N\right)\right]
\end{align}
Then the $(i,j)$-th element, $\forall i,j=1,2$, in FIM, i.e. $\mathbf{I}_N(\bm{\beta})_{i,j}$, is
\begin{align}
\mathbf{I}_N(\bm{\beta})_{1,1}&=\sum_{N=0}^n\left\{\left[\frac{ \partial \bar{f}(\lambda,\xi|N)}{ \partial \lambda}\right]^2\frac{1}{f(\lambda,\xi|N)}\right\}\label{I11_2}\\
\mathbf{I}_N(\bm{\beta})_{2,2}&=\sum_{N=0}^n\left\{\left[\frac{ \partial \bar{f}(\lambda,\xi|N)}{ \partial \xi}\right]^2\frac{1}{f(\lambda,\xi|N)}\right\}\label{I22_2}\\
\mathbf{I}_N(\bm{\beta})_{1,2}=\mathbf{I}_N(\bm{\beta})_{2,1}&=\sum_{N=0}^n\left[\frac{ \partial \bar{f}(\lambda,\xi|N)}{ \partial \lambda}\frac{ \partial \bar{f}(\lambda,\xi|N)}{ \partial \xi}\frac{1}{f(\lambda,\xi|N)}\right]\label{I21_2}
\end{align}

Equivalently, the FIM for $\bm{\theta}=(M,p)$ is
\begin{align}
    \mathbf{I}_N(\bm{\theta})_{1,1}=%&\int_{N\in \mathcal{N}}\left[\frac{ \partial^2f(M,p|N)}{ \partial M}\right]^2\frac{1}{f(M,p|N)}d N,\notag\\
    \sum_{N=0}^n\left\{\left[\frac{ \partial L(\bm{\theta}|N)}{ \partial M}\right]^2\frac{1}{L(\bm{\theta}|N)}\right\},\notag
    %=&\sum_{N=0}^n\left\{\left[f(M+1,p|N)-f(M,p|N)\right]^2\frac{1}{f(M,p|N)}\right\}.
    \label{eq:I11}
\end{align}
where $L(\bm{\theta}|N)$ is not differentiable again since it is discrete in $M$. We can find a approximated $f(M,p|N)$, i.e.  $\bar{f}(M,p|N)$, using the similar method in obtaining $\bar{f}(\lambda,\xi|N)$. Then we have
\begin{align}
&\frac{ \partial \bar{f}(M,p|N)}{ \partial M}\notag\\
=&\frac{\Gamma (M) p^N (1-p)^{M-N} }{N! (M-N)^2 \Gamma (M-N)}\left\{M (N-M) \left[\psi(M-N)-\psi(M)-\log (1-p)\right]-N\right\}
\end{align}

Similarly, we have 
\begin{align}
\mathbf{I}_N(\bm{\theta})_{2,2}=\sum_{N=0}^n\left\{\left[\frac{ \partial f(M,p|N)}{ \partial p}\right]^2\frac{1}{f(M,p|N)}\right\},\label{I22}
\end{align}
and
\begin{align}
\mathbf{I}_N(\bm{\theta})_{2,1}&=\mathbf{I}_N(\bm{\theta})_{1,2}\notag\\
&\approx\sum_{N=0}^n\left\{\frac{ \partial \bar{f}(M,p|N)}{ \partial M}\frac{ \partial f(M,p|N)}{ \partial p}\frac{1}{f(M,p|N)}\right).\label{I12}
\end{align}
where 
\begin{align}
\frac{ \partial f(M,p|N)}{ \partial p}=-\frac{M! p^{N-1} (1-p)^{M-N-1} (M p-N)}{N! (M-N)!}
\end{align}
The CRLB is given by the inverse of the FIM, %where $M_0$ and $p_0$ are the ground truth values
\begin{align}
\mathbf{C} = \mathbf{I}_N(\bm{\theta})^{-1}\big|_{M=M_0,p=p_0}.
\end{align}
Given that there are $\nu$ i.i.d. experiments,  so the underlying Cramer Rao lower bound  is
\begin{align}
\mathbf{C}_{\nu} = \frac{\mathbf{C}}{\nu} =  \frac{1}{\nu}\mathbf{I}_N(\bm{\theta})^{-1}\big|_{M=M_0,p=p_0}\label{eq:CRLB}
\end{align}

\begin{figure}
    \centering
    \includegraphics[width=.8\textwidth]{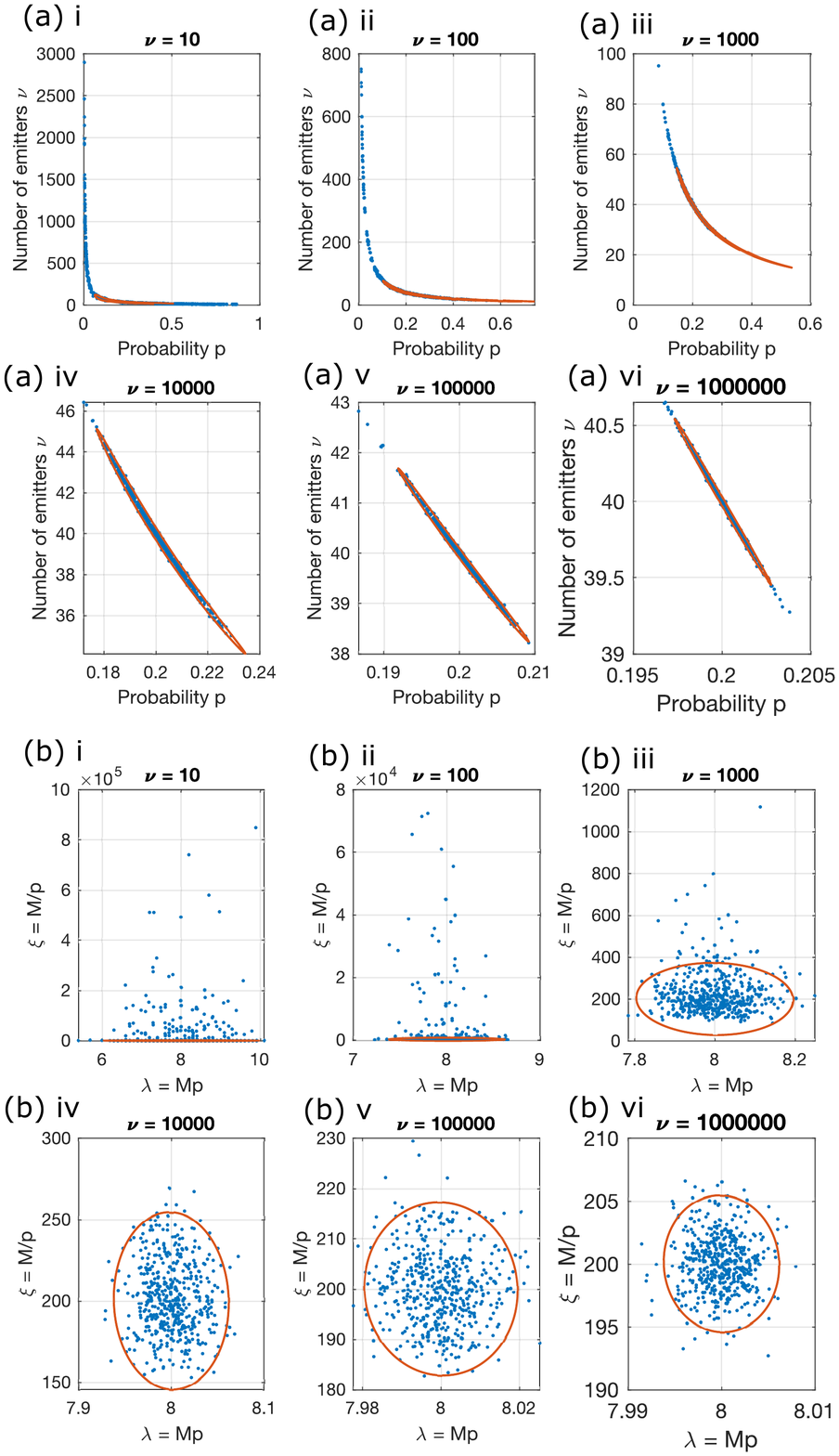}
    \caption{ The Monte-Carlo simulation results (blue dots) using maximum likelihood and the Cramer-Rao lower bound (red) showing as 95\% confidence interval ellipse in both (a) ($M,p$) space and (b) ($\lambda,\xi$) space. $p_0=0.2$ and $M_0=40$.}
    \label{fig:CRLB}
\end{figure}

We proceeded to compare our maximum likelihood simulations with the CRLB. The parameters $(M,p)$ are estimated using increasing number of experiments, i.e. $\nu$. For each $\nu$, we performed $500$ independent Monte Carlo simulations.  By performing an ensemble of numerical experiments, we could compare the estimated values in the $(M,p)$ space with a 2D confidence region, i.e. the CRLB 95\% error ellipse (within which the probability that the random estimated value $\bm{\theta}=(M,p)$ will fall inside the ellipse is 95\%). % determine the $95\%$ error ellipse, i.e. 2-d confidence region, which is then compared with the $\mathbf{C}_{\nu}$.
The simulation results are shown in Fig.~\ref{fig:CRLB} with ground truth $p_0=0.2$ and $M_0=40$. 

Fig.~\ref{fig:CRLB} shows a series of maximum likelihood determinations of the number of emitters and probability of detection per emitter, for ground truth $\bm{\theta_0} = (M,p) = (40,0.2)$.  We show the results in $\left(\lambda,\xi\right)$ and the data converted back into $\left(M,p\right)$ space, for increasing experiments $\nu$.  Each point represents the maximum likelihood determination and the solid curve is the 95\% confidence interval.  Observe that in $\left(\lambda,\xi\right)$ space we obtain a standard error ellipse in Fig.\ref{fig:CRLB} (b), whereas in $\left(M,p\right)$ space in Fig.\ref{fig:CRLB} (a), the ellipse is converted according to reciprocal functions: $p = \sqrt{\lambda/\xi}$ and $M = \sqrt{\lambda \xi}$. The shape of the error region in $\left(M,p\right)$ space is a consequence of the classical ambiguity between more dim emitters and fewer brighter emitters.  Nevertheless, as can be seen, by applying quantum measurements, some bounding on the number of emitters can be obtained, with increasing certainty as the number of experiments increases.  For simplicity, we have not enforced the requirement that CRLBs of $\left(M,p\right)$ and $\left(\lambda,\xi\right)$ are positive, although these values are strictly positive in the simulation data, hence the maximum likelihood and CRLB values do not agree for small $\nu$. 

\begin{figure}
    \centering
    \includegraphics[width=.8\textwidth]{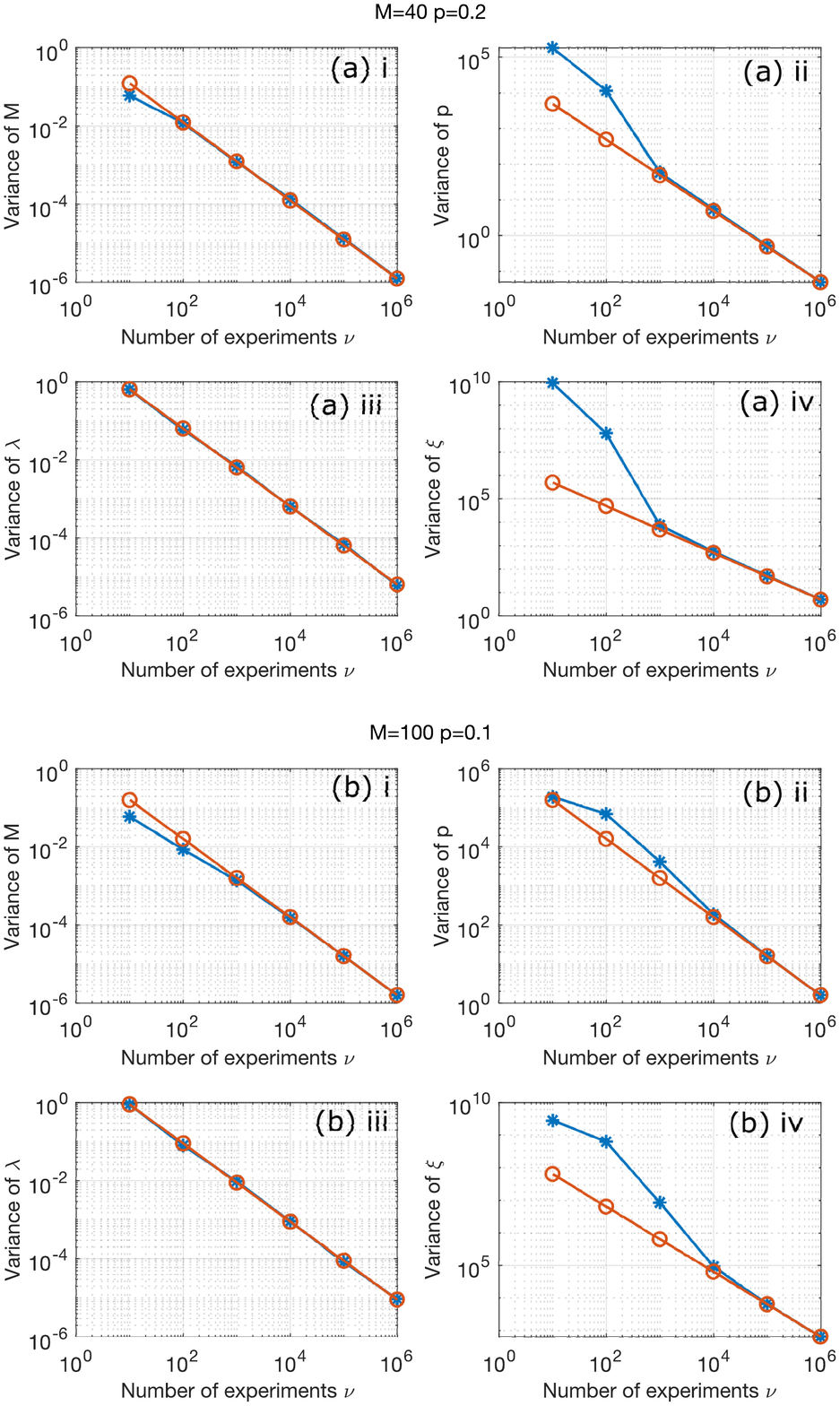}
    \caption{The scaling of Cramer-Rao lower bound (red) and the Monte-Carlo simulation (blue) using maximum likelihood. Ground truth (a) $p_0=0.2$, $M_0=40$ (b) $p_0=0.1$, $M_0=100$. (a)i-ii, (b)i-ii present variances in (M,P) space and (a)iii-iv, (b)iii-iv are variances in ($\lambda,\xi$) space. It can be observed that the sampled variance of $p$ is slightly smaller than the CRLB when the number of experiments is small. This is because the MLE is biased on those points due to the small size of data \cite{WANG201579}.}
    \label{fig:scalinglaw}
\end{figure}

To justify the performance of the maximum likelihood model, we compare the variances of predicted $M$ and $p$ results with CRLB. Fig.\ref{fig:scalinglaw} presents two configurations, $\bm{\theta_0} = \left(40,0.2\right)$ and $\bm{\theta_0} = \left(100,0.1\right)$ with CRLB in $(\lambda,\xi)$ space and $(M,P)$ space. Both showing asymptotic trend to CRLB. A log-log scaling law is observed here, i.e the $\log(\text{Variance})$ scales with $-\log(\nu)$. Ideally the variance of estimated data cannot be lower than CRLB, but for small $\nu$ the estimator is biased when there are too few measurement data, and CRLB only holds when the estimator is unbiased.

\begin{figure}
    \centering
    \includegraphics[trim=0cm 5cm 0cm 5cm, width=\textwidth]{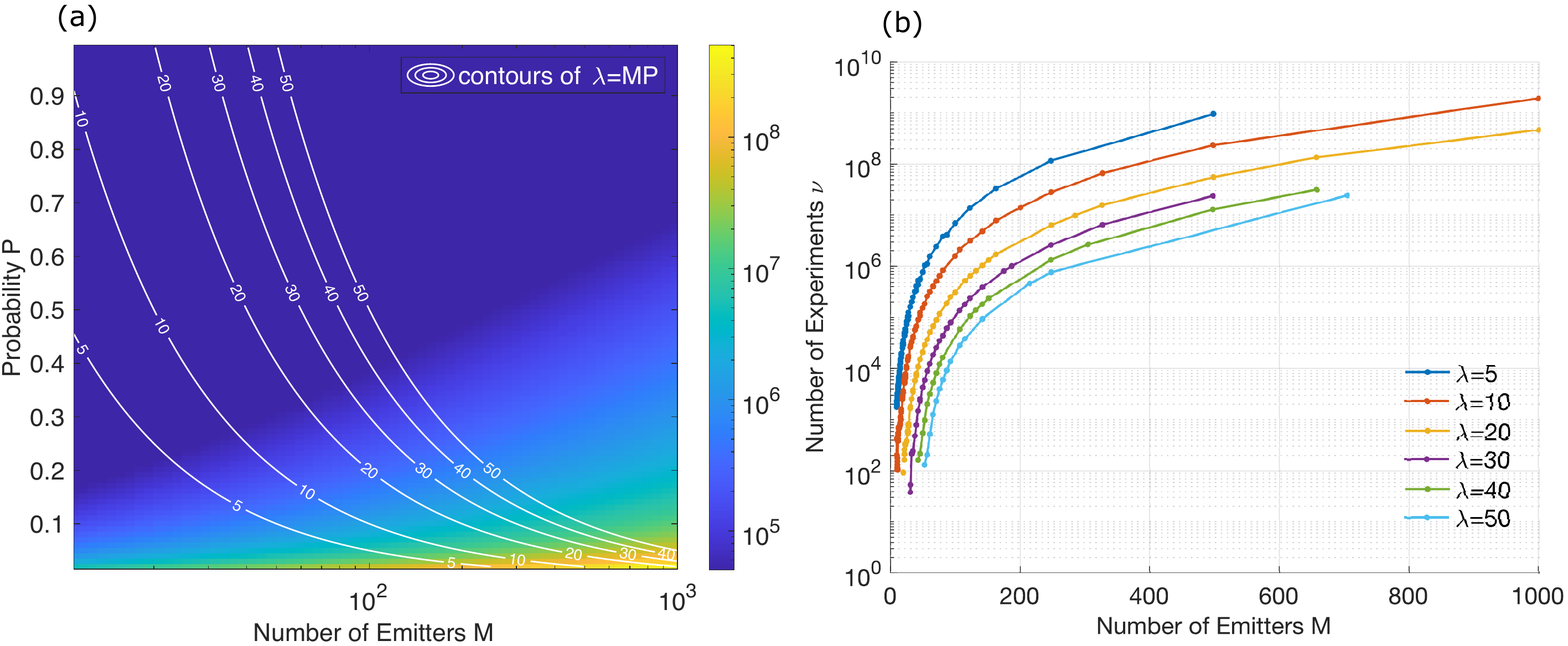}
    \caption{(a) Number of experiments to achieve CRLB with the relative variance of M: {$\mathrm{Var}[M]/M=1\%$}. Here the lower bound of variance of M is the CRLB(M) element in the matrix of Eq.\ref{eq:CRLB}. The contours show the $\lambda=MP$. (b) The relationships of $\nu$ and $M$ along contours with fixed $\lambda$ value, from $\lambda=5$ to 50, generated by extracting the $\nu$ values along contours in (a). }
    \label{fig:NexpMap}
\end{figure}

Fig.\ref{fig:NexpMap} (a) shows the number of experiments that required to meet the CRLB criterion, here we define the CRLB criterion as the relative variance of $M$: $\mathrm{Var}[M]/M=1\%$. Here the lower bound of variance of $M$ is the corresponding CRLB element in the matrix of Eq.~\ref{eq:CRLB}. The contours show the resolved photon number with max likelihood $\lambda=MP$. In the top half of the map where the probability $p$ roughly larger than 0.5, the number of experiments to achieve CRLB is relatively small ($<10^6$), even for large emitter numbers $10^3$. In the bottom half of the map where $p<0.5$, with the decrease of the probability $p$ the number of experiments to achieve CRLB increases dramatically. With such low brightness or detected probability when it reaches to large emitter numbers $M$, measurements that required to determine $M$ is several orders higher than high brightness scenario.
 
 Along one contour with a fixed $\lambda$, $\nu$ increases with the increase of $M$ and decrease of $p$, which means given a detected photon number distribution with the peak occurrence locating at $\lambda$ (similar to Fig.\ref{fig:BarFig}), more measurements are required to resolve many low brightness emitters than few high brighter emitters. Fig.\ref{fig:NexpMap} (b) shows the relationships of $\nu$ and $M$ along one contour with fixed $\lambda$, from $\lambda=5$ to 50. The curves show an elbow shape when approaching large amount of emitters. To the left of the elbow shape, measurements $\nu$ increases dramatically with emitter numbers. To the right of elbow shape, e.g.$M=200$, the small $\lambda$ curves stay on top of the large $\lambda$ ones. This is because the small $\lambda$ indicates a small probability of detecting photons from each emitter, and results in more measurements being required to determine the number of emitters.
 
We now consider the example of quantitative fluoresence. If we consider a sample of 1,000 fluorophores, in a field of view with probability of photon collection of $1\%$ from each emitter, then the number of measurements required to achieve a determination of the number of emitters with a relative variance of $1\%$ is around $1.96 \times 10^9$.  Photon number resolving measurements can be performed at of order microsecond timescales \cite{Morais2020arXiv}.  This means that the length of time required to achieve to determine the number of identical (but unknown) fluorescent emitters is if order $\sim$30min.  

%%%%%%%%%%%%%%%%%%%%%%%%%%%%%%%%%%%%%%%%%%%%%%%%%%%%%%%%%%%%%%%%
\section{Conclusions}
We have shown that photon number resolving measurements can help to identify the number of emitters in a field of view, even without \emph{a priori} knowledge of the brightness of the emitters.    Our results enable the prediction of the number of experiments required for a particular variance to be achieved.  As the number of emitters increases, the photon distribution approaches Poissonian, and in this limit, resolution of the number of emitters becomes increasingly difficult.

Our results show the idealised case, of equal brightness emitters.  Naturally, variations in the emission probability (for example by particles locate in different parts of the optical point spread function, or with different local environments) will lead to increased number of experiments.  Nevertheless, our analysis is likely to guide future experiments in quantitative tests of biological pathways.  Practical systems will also have to contend with variations in photo-bleaching of emitters that may limit the practically achievable number of experiments.  As such, our results  provide an opportunity to bound the expected sample variance, and hence to give limits on the number of emitters that might be contributing to a signal - bounds that are not possible to impose given the current limits of classical fluorescence based imaging.

%%%%%%%%%%%%%%%%%%%%% acknowledgements %%%%%%%%%%%%%%%%%%%%%%%%%%
\section*{Acknowledgements}
This work was funded by the Air Force Office of Scientific Research (FA9550-20-1-0276).  ADG also acknowledges funding from the Australian Research Council (CE140100003 and FT160100357).
VVY acknowledges partial support from the National Science Foundation (NSF) (DBI-1455671, ECCS-1509268, CMMI-1826078), the Air Force Office of Scientific Research (AFOSR) (FA9550-15-1-0517, FA9550-20-1-0366, FA9550-20-1-0367), Army Medical Research Grant (W81XWH2010777), the National Institutes of Health (NIH) (1R01GM127696-01, 1R21GM142107-01), the Cancer Prevention and Research Institute of Texas (CPRIT) (RP180588).

%%%%%%%%%%%%%%%%%%%%%%% references %%%%%%%%%%%%%%%%%%%%%%%%%
\section*{Reference}
\bibliography{main}

\end{document}